\documentstyle[12pt, fleqn]{article}

\newcommand{\rstr}{\hbox{
$\vert\mkern-4.8mu\hbox{\rm\`{}}\mkern-3mu$}} 

\newcommand{\QED}{\hfill{\vrule height 7pt width 7pt depth 0pt}\par\bigskip}

\title{Broken Symmetries in the Entanglement of Formation}

\author{Fabio Benatti\\
{\small Dip. Fisica Teorica, Universit\`a di Trieste}\\
{\small Strada Costiera 11, I-34100 Trieste, Italy}\\
{\small email: Benatti@Trieste.infn.it}\\
H. Narnhofer\\
{\small Inst. f\"{u}r Theoretische. Physik, Vienna University, Austria}\\
{\small email: narnh@ap.univie.ac.at}\\
A. Uhlmann\\
{\small Inst. f\"{u}r Theoretische Physik, Leipzig University, Germany}\\
{\small email: Armin.Uhlmann@itp.uni-leipzig.de}
\\
}

\date{\null}

\begin{document}

\maketitle

\begin{abstract}
We compare some recent computations of the
entanglement of formation in quantum information theory and of the
entropy of a subalgebra in quantum ergodic theory.
Both notions require optimization over decompositions of quantum states.
We show that both functionals are strongly related for some
highly symmetric density matrices. Indeed,
for certain interesting regions the entanglement
of formation  can be expressed by the entropy of a commuting
subalgebra, and the corresponding optimal decompositions
can be obtained one from the other.
We discuss the presence of broken symmetries in relation
with the structure of the optimal decompositions.
\end{abstract}

\noindent
{\bf 1. INTRODUCTION}
\medskip

Entanglement, always one of the most intriguing among quantum marvels,
has lately become a powerful resource in prospective quantum
information technologies~\cite{NC};
measuring the entanglement content of states of multipartite quantum
systems is thus of great practical importance.
If a
bipartite system $A+B$ is described by a density matrix
$\rho_{AB}$,
the so-called entanglement of formation~\cite{BVSW} is measured by
\begin{equation}
\label{entform}
E(\rho_{{AB}}):=\inf\Bigl\{
\sum_j \lambda_jS({\rm Tr}_B\pi_j)\, :\,
\rho_{AB}=\sum_j \lambda_j\pi_j\Bigr\}\ .
\end{equation}
In the above expression, $S(\rho):=-{\rm Tr}\rho\log\rho$
denotes the von Neumann entropy of the state obtained by partial trace
over $B$ and
the infimum is computed over all
possible decompositions of
$\rho$ as convexly linear
combinations, that is $\lambda_j>0$, $\sum\lambda_j=1$,
of one-dimensional projections $\pi_j$ of $A+B$.
In the following we call such decompositions
{\it extremal convex decompositions of $\rho$}
to be distinguished
from generic convex decompositions into mixed states.

When $\rho_{AB}=|\Psi_{AB}\rangle\langle\Psi_{AB}|$,
the entanglement of formation gives the asymptotic ratio between
the number of singlet states necessary to construct $N\gg 1$ copies of
$\rho_{AB}$~\cite{BBPS}.
In the following, we will compare the entanglement of formation
with a particular case of a
more general quantity, the ``entanglement with
respect to a subalgebra'' or ``entanglement'', for short. This latter
concept is related to the so-called ``entropy of a subalgebra''
${\cal A}$ contained in a reference algebra ${\cal M}$, relative
to a state $\rho$ on ${\cal M}$~\cite{NT},
\begin{equation}
\label{entsub}
H_\rho({\cal A}):=S(\rho\rstr{\cal A})\,-\,
\inf\Bigl\{\sum_j \lambda_jS(\rho_j\rstr{\cal A})\,:\,
\rho=\sum_j\lambda_j\rho_j\Bigr\}\ .
\end{equation}
In the above expression,
the infimum is calculated over all convexly linear decompositions
of $\rho$ into other states on $\cal M$.
It plays a key role in extending
the classical dynamical entropy of Kolmogorov to quantum
systems~\cite{CNT,B,OP}.
The entanglement of formation~(\ref{entform}) can be considered
a special case of~(\ref{entsub}).

We shall call ``optimal'' those decompositions achieving the extremum
in~(\ref{entform}) and~(\ref{entsub}).
Calculating either $E(\rho_{AB})$ or
$H_{\rho}({\cal A})$ is particularly complicated.
The problem has been completely solved for the entanglement of
formation if
${\cal H}_A={\cal H}_B={\bf C}^2$~\cite{HW}, and
for the entropy of a subalgebra if
${\cal M}=M_2({\bf C})$~\cite{Le94,BG,U}.
So far, all other available results concern
states $\rho_{AB}$ and $\rho$ that are highly
symmetric, isotropic in~\cite{TV}, respectively permutation-invariant
in~\cite{BNU1}.

In this paper we will discuss the previously mentioned results by
comparing the two notions of entanglement sketched above. We
show, that some of them are one-to-one related.
To do so, we shall
focus on the structure of optimal decompositions in relation to
the symmetries existing in the problem and show possible ways
of breaking them.
These symmetries form a group $G$ and leave invariant both
the state $\rho$ and, as a set, the subalgebra $\cal A$.
Given extremal optimal decompositions, the $G$-orbits of each
of their pure states consist of optimal decomposers, too.
We will study the dependence of either entanglements upon the
number of different orbits.
\medskip

\noindent
{\bf 2. ENTANGLEMENT}
\medskip

In the following, we shall consider quantum systems described by
algebras of operators, $\cal M$,
acting on finite or infinite dimensional
Hilbert spaces $\cal H$, with states,
${\cal M}\ni X\longmapsto{\rm Tr}(\rho\, X)$,
represented by density matrices which we shall denote
by greek letters.
\medskip

\noindent
{\bf Definition 2.1}\quad
Given a finite dimensional subalgebra ${\cal A}\subseteq{\cal M}$,
we define the entanglement of the state $\rho$ with respect to $\cal A$ by
\begin{equation}
E\Bigl(\rho;{\cal M},{\cal A}\Bigr):=\inf\Bigl\{
\sum_j \lambda_jS(\rho_j\rstr{\cal A})\,:\,
\rho=\sum_j \lambda_j\rho_j\Bigr\}\ ,
\label{ent}
\end{equation}
where $\rho=\sum_j \lambda_j\rho_j$ runs through all convexly linear
decompositions of $\rho$ with
states of $\cal M$, and
$S(\rho_j\rstr{\cal A})$ is
the von Neumann entropy of the state $\rho_j$ restricted to  the
subalgebra $\cal A$. The entanglement
(\ref{ent}) is convex as a function of $\rho$.
\medskip

\noindent
{\bf Remarks 2.1}

\noindent
$(i)$\quad
The entanglement~(\ref{ent}) is a convex functional over the states:
\begin{equation}
\label{convent}
E\Bigl(\sum_j\mu_j\rho_j;{\cal M},{\cal A}\Bigr)\leq
\sum_j\mu_jE\Bigl(\rho_j;{\cal M},{\cal A}\Bigr)\ ,\quad
\sum_j\mu_j=1\ ,\ \mu_j\geq0\ .
\end{equation}
This follows by choosing optimal decompositions for the $\rho_j$'s,
which together provide a decomposition, not necessarily optimal, for 
$\sum_j\mu_j\rho_j$.

\noindent
$(ii)$\quad
The entanglement of formation~(\ref{entform})
is the entanglement~(\ref{ent}) with $\cal A$, respectively $\cal B$, 
the algebra of observables of the system $A$, respectively $B$,
${\cal M}={\cal A}\otimes{\cal B}$ and
$\rho_{AB}\rstr{\cal A}={\rm Tr}_B\rho_{AB}$.

\noindent
$(iii)$\quad
The entanglement (\ref{ent}) is related with
the entropy of a subalgebra~(\ref{entsub}) by
\begin{equation}
\label{entsub1}
E(\rho_{AB})=S\left(\rho_{AB}\rstr{\cal A}\otimes{\bf 1}_B\right)\,
-\,H_{\rho_{AB}}\left({\cal A}\otimes{\bf 1}_B\right)\ .
\end{equation}
Indeed, as we shall see below in Proposition 2.1, the infimum is
achieved at decompositions using pure states of $\cal M$ only, and
it enjoys some further remarkable properties.
\medskip

The quantity in (\ref{entsub1}) and some techniques~\cite{BNU1,BNU2} that
were developed for
computing~(\ref{entsub}),
have recently been used to attack the question whether the entanglement of
formation is additive~\cite{BN}.
Among them, a useful result is contained in the following
proposition. The idea is in~\cite{BNU1} and, slightly extended,
in \cite{Uh98}.
We include a proof for the benefit of the reader.
\smallskip

\noindent
{\bf Proposition 2.1}\quad
If the algebra $\cal M$ is finite dimensional then
\begin{itemize}
\item
the entanglement $E\Bigl(\rho;{\cal M},{\cal A}\Bigr)$ is achieved at
certain extremal convex decompositions
$\rho=\sum_j \lambda_j \pi_j$, $\lambda_j >0$
which saturate (\ref{ent}).
Such decompositions
are called {\em optimal.} Every pure state, $\pi$,
which appears in at least one optimal decomposition
of $\rho$  is called
{\em $\rho$-optimal} or an {\em optimal decomposers}
of $\rho.$
\item
For every $\rho$ there is an optimal decomposition
with a length not exceeding the linear dimension
of $\cal M$.
\item
The functional $E\Bigl( \, . \, ;{\cal M},{\cal A}\Bigr)$
is convexly linear on the convex hull
${\cal R}(\rho)$ of all $\rho$-optimal pure states:
Let be
$\omega =\sum_i \alpha_i \pi_i$, $\alpha_i > 0$,
$\sum_i \alpha_i=1$ any extremal convex decomposition where
the $\pi_j$ are some optimal
decomposers of $\rho$. Then
\begin{equation}
\label{subdec}
E\Bigl(\omega;{\cal M},{\cal A}\Bigr) =\sum_i
\alpha_i S(\pi_i\rstr{\cal A})\ .
\end{equation}
\end{itemize}
\smallskip

\noindent {\bf Proof:}\quad Any mixed state $\rho$ can be
decomposed and, since the von Neumann entropy is concave on convex
combinations, mixed states cannot improve (\ref{ent}) with respect
to pure states. If $\cal M$ is $d$ dimensional, compactness of the
state space, extremality and compactness of the set of pure states
ensure by a theorem of  Caratheodory that we need not less than
$d$ and not more than $d^2$ decomposers~\cite{U,R}. Because of
convexity (\ref{convent}), the functional
$E\Bigl( \, . \, ;{\cal M},{\cal A}\Bigr)$ is the supremum over affine
functionals. Thus, for every $\rho$ there are functionals
$\ell$ such that $E\Bigl(\rho;{\cal M},{\cal A}\Bigr)=
\ell(\rho)$, while, on generic states $\sigma$,
$E\Bigl(\sigma;{\cal M},{\cal A}\Bigr)\geq \ell(\sigma)$. Given an
optimal decomposition $\rho=\sum_j \lambda_j\pi_j$ it follows
\begin{eqnarray}
\nonumber
E\Bigl(\rho;{\cal M},{\cal A}\Bigr) &=&
\sum_j \lambda_j E\Bigl(\pi_j;{\cal M},{\cal A}\Bigr)\\
&\geq&
\sum_j\lambda_j \ell(\pi_j)= \ell(\rho) =
E\Bigl(\rho;{\cal M},{\cal A}\Bigr)\ .
\label{th211a}
\end{eqnarray}
Since equality must hold in (\ref{th211a}) and because
$\lambda_j > 0$, while 
$E\Bigl(\pi_j;{\cal M},{\cal A}\Bigr) \geq \ell(\rho)$
by assumption, we conclude
$E\Bigl(\pi_j;{\cal M},{\cal A}\Bigr)=\ell(\pi_j)$ for all $j$.
With $\omega \in {\cal R}(\rho)$, let us now fix this affine functional
$\ell$ and consider the extremal decomposition
$\omega = \sum \alpha_k \pi'_k$ such that all the $\pi'_i$
are optimal decomposers of $\rho$. By convexity and the preceding
argument we deduce
\begin{equation} \label{th213a}
E\Bigl(\omega;{\cal M},{\cal A}\Bigr)  \leq
\sum_k \alpha_k E\Bigl(\pi'_k;{\cal M},{\cal A} \Bigr) =
\sum_k \alpha_k \ell(\pi'_k) = \ell(\omega)
\end{equation}
However,
$ \ell(\omega) \leq E\Bigl(\omega;{\cal M},{\cal A}\Bigr)$
by our choice of $\ell$, and equality holds in (\ref{th213a}).
Thus, $E\Bigl(\cdot;{\cal M},{\cal A}\Bigr)$
is convexly linear on ${\cal R}(\rho)$.
\QED

\noindent
{\bf Definition 2.2}\quad
We shall call the convex hull ${\cal R}(\rho)$
of the optimal decomposers of $\rho$ a
{\it leaf} with respect to the entanglement
$E\Bigl(\rho;{\cal M},{\cal A}\Bigr)$.
Then, the state space appears as covered by leaves, and
the entanglement itself is convexly linear above every leaf.
That effect is referred to as the {\it roof property}
of $E\Bigl(\,\cdot\,;{\cal M},{\cal A}\Bigr)$,
\cite{U}, i.e.
$E\Bigl(\,\cdot\,;{\cal M},{\cal A}\Bigr)$ is a convex roof.
\medskip

\noindent
{\bf Definition 2.3}\quad Given $\rho$ on $\cal M$, we
shall call a group $G$ a symmetry group with respect to
$E(\rho;{\cal M},{\cal A})$, if for all $g\in G$ there exists a
linear map $\gamma_g:{\cal M}\longmapsto{\cal M}$ such that the
state and the subalgebra $\cal A$ (as a set) are left invariant
by $\gamma_g$, Namely, $\gamma^*_g[\rho]=\rho$, where
$\gamma^*_g[\rho](m)={\rm Tr}(\rho\gamma_g(m))$.
\medskip

\noindent
{\bf Proposition 2.2}\quad
If $G$ is a symmetry group with respect to
$E(\rho;{\cal M},{\cal A})$,
the leaf ${\cal R}(\rho)$ is $G$-invariant as a set.
In particular, the action of $G$ permutes the optimal decomposers of
$\rho$.
\medskip

\noindent
{\bf Proof:}\quad
Let $\rho=\sum_{j\in J}\lambda_j\rho_j$ be an optimal decomposition 
with respect to $E(\rho;{\cal M},{\cal A})$.
Then, since $\gamma_g^*[\rho]=\rho$ and 
$\gamma({\cal A})={\cal A}$ for $g\in G$, the decomposition
$\rho=\sum_{j\in J}\lambda_j\gamma^*_g(\rho_j)$ is also optimal.
Therefore, its leaf ${\cal R}(\rho)$ must contain both the $\rho_j$'s
and the $\gamma^*_g(\rho_j)$'s.
\QED

Based on the previous two propositions, the entropy $H_\rho({\cal A})$ has
explicitly been computed in the following cases,
\smallskip

\noindent
{\bf Case 1.}~\cite{Le94,BG,U}\quad
Let ${\cal M}$ be the full $2\times 2$
matrix algebra $M_2({\bf C})$, ${\cal A}$ the subalgebra of all
$2\times 2$ matrices diagonal with respect to a given basis
$|1\rangle$, $|2\rangle$, and
$\displaystyle
\rho=\pmatrix{a&b\cr b^*&1-a}$, $0\leq a\leq 1$, $|b|^2\leq a(1-a)$, any
density matrix.
\medskip

\noindent
{\bf Case 2.}~\cite{BNU1}\quad
Let ${\cal M}=M_3({\bf C})$, ${\cal A}$ the subalgebra of all
$3\times 3$ diagonal matrices with respect to the basis
$|1\rangle$, $|2\rangle$, $|3\rangle$ and
\begin{equation}
\label{3perminv}
\rho(x)={1\over 3}\pmatrix{1&x&x\cr x&1&x\cr x&x&1}\ ,\quad
-1/2\leq x\leq 1\ ,
\end{equation}
any density matrix invariant under the group of
permutations of $\{1,2,3\}$.
\medskip

For future comparison with the entanglement of formation
of isotropic states of
$d$-dimensional bipartite systems studied in~\cite{TV},
we fix an orthonormal basis $|j\rangle\in{\bf C}^d$
and consider the group $G$ of permutations of $\{1,2,\ldots, d\}$.
It turns out that any $G$-invariant density matrix
$\rho(x)$ over ${\cal M}=M_d({\bf C})$ can be written as
\begin{equation}
\label{F1}
\rho_F={1-F\over d-1}\Bigl({\bf 1}-|\psi\rangle\langle\psi|\Bigr)\,+\,
F|\psi\rangle\langle\psi|\ ,
\end{equation}
where $\displaystyle
|\psi\rangle={1\over\sqrt{d}}\sum_{j=1}^d|j\rangle$
and $F$ is the fidelity parameter
\begin{equation}
\label{F2}
0\leq F:=\langle\psi|\rho(x)|\psi\rangle
={(d-1)x+1\over d-1}\leq 1\ .
\end{equation}
Setting $s(t):=-t\log t$, we have,

\noindent
{\bf Case 1.}\quad
For all $\rho$, the optimal decompositions are
\begin{eqnarray}
\label{opdec2a}
\rho&=&\lambda|w_1\rangle\langle w_1|+(1-\lambda)|w_2\rangle\langle
w_2|\\
\label{opdecb}
|w_1\rangle&=&\pmatrix{z_1\cr z_2}\ ,|w_2\rangle=\pmatrix{z_2^*\cr
z_1^*}, b = z_1 z_2^*,\\
\label{opdecc}
|z_1|^2&=&(1+\sqrt{1-4|b|^2})/2=1-|z_2|^2\ ,\
\lambda={1\over 2}\left(1+{2a-1\over\sqrt{1-4|b|^2}}\right)\ .
\end{eqnarray}
The corresponding entanglement is
$E(\rho;M_2({\bf C}),{\cal A})=s(|z_1|^2)+s(|z_2|^2)$.

\noindent
If $\rho=\rho_F$ is permutation-invariant, that is, if $a=1/2$, $b=x/2$
$F=(1+x)/2$, the entanglement reads
\begin{equation}
\label{opdec2b}
E(\rho_F;M_2({\bf C}),{\cal A})=s\Bigl({1+2\sqrt{F(1-F)}\over 2}\Bigr)+
s\Bigl({1-2\sqrt{F(1-F)}\over 2}\Bigr)\ .
\end{equation}
\smallskip

\noindent
{\bf Case 2.}
Given the group $G$ of permutations of $\{1,2,3\}$,
let $V$, $V^2$ implement unitarily the subgroup $G_0$ of
cyclic permutations.
Then, any $G$-invariant state $\rho_F$
can be written
\begin{equation}
\label{opdecsym}
\rho_F={1\over 3}|w\rangle\langle w|+
{1\over 3}V|w\rangle\langle w|V^{-1}+
{1\over 3}V^2|w\rangle\langle w|V^{-2}\ ,
\end{equation}
where 
\begin{equation}
\label{opdec4}
|w\rangle={1\over 3}\pmatrix{
a+2b\cos\theta\cr a-2b\cos(\theta-\pi/3)\cr a-2b\cos(\theta+\pi/3)}\ ,
\quad
a=\sqrt{3 F}\ ,\quad
b=\sqrt{{3\over 2}(1-F)}\ .
\end{equation}
The structure of optimal decompositions depends on the convexity of
\begin{equation}
{\cal S}(F):=\min_{\theta\in[0,2\pi]}\sum_{j=1}^3s(|w_j(F;\theta)|^{2})\ .
\label{conc}
\end{equation}
For $F\geq F^*:=(2x^*+1)/3$, $x^*=-0.4150234$, the minimum is
achieved at a single extremal
$G_0$-orbit generated by the vectors
\begin{equation}
\label{opdec5}
|w\rangle={1\over 3}\pmatrix{a+2b\cr a-b\cr a-b} =
{1 \over \sqrt{3}}
\pmatrix{ \sqrt{F} + \sqrt{2(1-F)} \cr \sqrt{F} - \sqrt{(1-F)/2} \cr
\sqrt{F} - \sqrt{(1-F)/2} }
\end{equation}
For each $0 < F < F^*$, there are two different orbit-generating vectors,
$|w_\pm(F)\rangle$, whose $G_0$-orbits provide different optimal
decomposers for (\ref{conc}), and which form together one orbit
of the full permutation group $G$.   They are
\begin{equation}
\label{opdec6}
|w_\pm(F)\rangle={1\over 3}\pmatrix{a+2b\cos\alpha_F\cr
a-2b\cos(\pi/3\mp\alpha_F)\cr
a-2b\cos(\pi/3\pm\alpha_F)}\ ,
\end{equation}
where the angle $\alpha_F$ varies with $0<F < F^*$.

Finally, for $F=0$, $\alpha_F=-\pi/6$,
the minimum is achieved again at a single $G$-orbit containing
the vector, $\displaystyle |w_0\rangle={1\over\sqrt{2}}(1,0,-1)$.
As the 6 vectors coincide pairwise up to a sign, the states
form a single optimal decomposition of length 3.

In \cite{BNU1}), it is shown that the above vectors give
optimal decompositions as long the function ${\cal S}(F)$ is convex.
Numerically, this is the case for all $F \leq 8/9$.
The corresponding entanglement is
\begin{eqnarray}
\nonumber
E(\rho_F;M_3({\bf C}),{\cal A})&=&s\Bigl({2-F+2\sqrt{2F(1-F)}\over3}\Bigr)\\
\label{opdec8}
&+&2\,s\Bigl({1+F-2\sqrt{2F(1-F)}\over 6}\Bigr)\ .
\end{eqnarray}
for fidelities $F^* \leq F \leq 8/9$. For $F=0$ the entanglement
equals $\log 2$. We have only numerical results within
the interval $0 < F < F^*$, \cite{BNU2}, reflecting that the
exact dependence of the angle
$\alpha_F$ in (\ref{opdec6}) as a function of $F$ is unknown.

\noindent
{\bf Remark 2.2}\quad
Permutation-invariant states as in~(\ref{F1}) can be written as
averages over the unitaries $U_\pi$ implementing the permutation
group $G$,
\begin{equation}
\label{ergav}
\rho_F={1\over d!}\sum_{\pi}U^{-1}_\pi|\phi\rangle\langle\phi|U_\pi\ ,
\end{equation}
if and only if $|\langle\psi|\phi\rangle|^2=F$, where $|\psi\rangle$
is the vector in~(\ref{F2}).
Necessity comes from the fact that $U_\pi|\psi\rangle=|\psi\rangle$.
Sufficiency: The identity ${\bf 1}$ and
$|\psi\rangle\langle\psi|$ form a basis for all possible contributions
to the averages~(\ref{ergav}).
\medskip

In view of the structure of the optimal decomposers discussed
above, we introduce a notion of regularity with respect to a
subgroup of a symmetry group, as follows.
\smallskip

\noindent
{\bf Definition 2.4}\quad
Given a symmetry group $G$ with respect to
$E(\rho;{\cal M},{\cal A})$,
we shall call a leaf ${\cal R}(\rho)$ regular of order $n$ with respect to a
subgroup $H\subset G$, if there exist $n$ pure states
$\bar{\rho}_j\in {\cal R}(\rho)$ such that $\gamma_h^*[\bar{\rho}_j]
=\bar{\rho}_j$ for all $h\in H$, whereas the convex span of the orbits
$\Bigl\{\gamma_g^*[\bar{\rho}_j]\Bigr\}_{g\in G}$ is the whole of
${\cal R}(\rho)$.
\medskip

\noindent
We illustrate the previous definitions with some examples.
\smallskip

\noindent
{\bf Example 2.1}\quad
Let $\cal M$ be a full $d\times d$ matrix algebra on ${\bf C}^d$ and
${\cal A}\subset {\cal M}$ diagonal with respect
to a chosen orthonormal basis $\{|j\rangle\}_{j=1}^d$ in ${\bf C}^d$.
Let $\rho$ be a symmetric density matrix, $\langle j|\rho|k\rangle=\langle
k|\rho|j\rangle$.
Then, with respect to the chosen representation,
the transposition $\cal T$ respects both the state and 
the subalgebra $\cal A$.
Also, ${\cal R}(\rho)$ is regular with respect to
$G=H=\{{\rm id},{\cal T}\}$, the order of regularity depending on the
state $\rho$.
In fact, let $\pi=|\psi\rangle\langle\psi|\in{\cal R}(\rho)$, then,
because of Proposition 2.2, 
${\cal T}(\pi)=\pi'=|\psi'\rangle\langle\psi'|\in{\cal R}(\rho)$, 
too. If $\pi\neq\pi'$, we may consider the state
$\omega=\pi/2+\pi'/2$. which, by Proposition 2.1, is already optimally 
decomposed. Also, 
\begin{equation}
\label{proof1}
E(\omega;{\cal M},{\cal A})=S(\pi\rstr{\cal A})=S(\omega\rstr{\cal
A})\ .
\end{equation}
Instead, the decomposition
\begin{eqnarray}
\label{proof2a}
\omega&=&{1+{\cal R}e(\langle\psi|\psi'\rangle)\over 2}\pi_+\,+\,
{1-{\cal R}e(\langle\psi|\psi'\rangle)\over 2}\pi_-\ ,\quad
\hbox{where}\\
\label{proof2b}
\pi_\pm&=&{|\psi\pm\psi'\rangle\langle\psi\pm\psi'|
\over 2(1\pm{\cal R}e(\langle\psi|\psi'\rangle)}
\end{eqnarray}
need not be optimal.
However, the concavity of the von Neumann entropy yields
\begin{eqnarray}
\nonumber
E(\omega;{\cal M},{\cal A})&\leq&
{1+{\cal R}e(\langle\psi|\psi'\rangle)\over 2}S(\pi_+\rstr{\cal A})\\
\label{proof3}
&+&{1-{\cal R}e(\langle\psi|\psi'\rangle)\over 2}S(\pi_-\rstr{\cal A})
\leq(S(\omega\rstr{\cal A})\ .
\end{eqnarray}
It thus follows from (\ref{proof1}) that 
$\pi\rstr{\cal A}=\pi_\pm\rstr{\cal A}$, whence 
the components $\psi(i)$, $\psi'(i)$ of $\psi$ and $\psi'$ 
must coincide apart from an overall phase.
Thus, $\pi=\pi'$ and the
$\cal T$-symmetry cannot be broken.
\smallskip

\noindent {\bf Example 2.2}\quad
Let ${\cal M}={\cal A}\otimes{\cal B}$, with $\cal A$ and $\cal B$
isomorphic and
$\sigma:{\cal A}\longmapsto{\cal B}$ the algebraic exchange of the
two of them. If $\rho$ is a state on $\cal M$ such that
$\rho\circ(\sigma^{-1}\otimes\sigma)=\rho$, in general,
$\sigma^{{-1}}\otimes\sigma$ does not belong to any subgroup of
regularity of $\rho$; indeed, if $\cal A$ (and thus $\cal B$) is a
$d$-dimensional matrix algebra and $\{|\ell\rangle\}$ is an
orthonormal basis in the corresponding Hilbert space ${\cal H}_A$
(and thus also in ${\cal H}_B$), the density matrix
\begin{equation}
\rho_{AB}:={1\over 2}|1\rangle\langle1|\otimes|2\rangle\langle2|\,
+\,{1\over 2}|2\rangle\langle2|\otimes|1\rangle\langle1|\ ,
\label{swap}
\end{equation}
is such that
${\rm Tr}\Bigl(\rho (\sigma^{-1}\otimes\sigma)(X\otimes Y)
\Bigr)={\rm Tr}\Bigl(\rho (X\otimes Y)\Bigr)$.
Also, $\rho_{AB}$ is already
optimally decomposed, $E(\rho_{AB};{\cal A},{\cal M})=0$ is
achieved with the decomposers $|1\rangle\langle
1|\otimes|2\rangle\langle2|$ and $|2\rangle\langle
2|\otimes|1\rangle\langle1|$, which, however, are not invariant
under $\sigma^{-1}\otimes\sigma$.
\smallskip

\noindent {\bf Example 2.3}\quad
Let ${\cal M}={\cal A}\otimes{\cal B}$, with $\cal A$ and $\cal B$
both $d\times d$
full matrix algebras. We fix the same orthonormal basis
$\{|\ell\rangle\}$ in both Hilbert spaces ${\cal H}_{A,B}$ and
consider the one-parameter group $U$ of unitaries
\begin{equation}
\label{unitgroup}
U_t:=\sum_{j,k}{\rm e}^{it(h_j-h_k)}\,
|j\rangle\langle j|\otimes|k\rangle\langle k|\ .
\end{equation}
The density matrix $\rho_{AB}:=\sum_{j,k}R_{jk}\,|j\rangle\langle
k|\otimes|j\rangle\langle k|$, \ , $R=[R_{jk}]\geq 0$,
${\rm Tr}R=1$, is $U$-invariant; moreover,
$\sqrt{\rho_{AB}}=\sum_{j,k}(\sqrt{R})_{jk}|\,
|j\rangle\langle k|\otimes|j\rangle\langle k|$, so that the operators
$\sqrt{\rho_{AB}}M\sqrt{\rho_{AB}}$, $M\in {\cal M}$, have the
same matrix structure as $\rho_{AB}$. Choosing positive
$M_j\geq 0$, $j\in
J$, such that $\sum_{j\in J}M_j=1$, $\rho_{AB}$ decomposes
into
\begin{equation}
\label{dec2} \rho_{AB}=\sum_{j\in J}\Bigl({\rm Tr}(\rho_{AB}\,
M_j)\Bigr)\, {\sqrt{\rho_{AB}}M_j\sqrt{\rho_{AB}}\over{\rm
Tr}(\rho_{AB}\,M_j)}\ .
\end{equation}
Since it is also true that every mixed state $\rho$ on $\cal M$
can be written
as in (\ref{dec2}) by means of a suitable positive $M_j$,
(\ref{dec2}) indeed exhausts all possible decompositions of
$\rho_{AB}$.
Thus, the decomposers $\pi_j$ of $\rho_{AB}$ which are optimal
with respect to $E(\rho_{AB};{\cal M},{\cal A})$, have
the same structure of $\rho_{AB}$ and are then $U$-invariant.
Hence, the group $U$ is a group of symmetries of $\rho_{AB}$
with respect to entanglement and the leaf
${\cal R}(\rho_{AB})$ is regular with respect to $H\equiv U$,
its order depending on which further symmetries are enjoyed
by $\rho_{AB}$.
\smallskip

\noindent {\bf Example 2.4}\quad Let ${\cal M}=M_2({\bf C}),
\cal{A}$
as in Case $1$, and $\rho_F$ a permutation-invariant state. The
leaf ${\cal R}(\rho_F)$ is the orbit of the group $G$ of
permutations of $\{1,2\}$. This follows from the form
of the optimal vectors ~(\ref{opdec2a}) in such a case:
$|w_1\rangle=\pmatrix{z_1\cr z_2}$, $|w_2\rangle=\pmatrix{z_2\cr
z_1}$, with $z_{1,2}=\sqrt{1/2(1\pm2\sqrt{F(1-F))}}$. It is regular
of order $1$ with respect to rotations with elements from ${\cal A}$.
\smallskip

\noindent
{\bf Example 2.5}\quad
Let ${\cal M}=M_d({\bf C})$ and $\rho_F$ a permutation-invariant state.
Then, for $F^*\leq F$ and $F$ belonging to the convexity region
of ${\cal S}(F)$
in~(\ref{conc}), the structure of the optimal vectors~(\ref{opdec5})
ensures that the leaf ${\cal R}(\rho_F)$ is regular of order $1$ for
the subgroup $H$ of permutations $\{2,3\}\mapsto\{3,2\}$.
However, at the point $F=F^*$ such a $H$-invariant
vector bifurcates into the two optimal ones~(\ref{opdec6}).
Thus regularity with respect to the subgroup $H$
is broken and remains broken for $0 < F < F^*$. At
$F=0$ optimal vector states of different $G_0$ orbits
degenerate pairwise into a single one, and one of them is
$H$-invariant, while the corresponding vector changes its sign.
\medskip

In the last two examples, for all $F$ when $d=2$, and
for $F$ greater than the bifurcation values $F^*$ in the
convexity region of ${\cal S}(F)$ in~(\ref{conc}), when $d=3$,
the leaf ${\cal R}(\rho_F)$ of a permutation-invariant $\rho_F$ is
generated by the orbit under the subgroup $G_0$ of
cyclic permutations
$V^j|w\rangle$, $j=0,1,2$. The vector $|w\rangle$ is
invariant under a unique transposition out of $G$.
This structure is indeed more
general as will be showed in the next two propositions.
\smallskip

\noindent {\bf Proposition 2.3}\quad
Let ${\cal A}\subset{\cal M}=M_d({\bf C})$ be chosen as in Example 2.1
and the density matrix $\rho_F$ be invariant with respect to the permutation
group $G$.
If the leaf ${\cal R}(\rho_F)$ with respect to $\cal A$ is generated by
exactly one
$G_0$-orbit of a normalized
vector state $|w\rangle\in{\bf C}^d$, with $G_0\subset G$ the subgroup of
cyclic permutations, then the entanglement is
\begin{eqnarray}
\label{entd0}
E(\rho_F;M_d({\bf C},{\cal A})&=&
s(p_F)+(d-1)s\Bigl({1-p_F\over d-1}\Bigr)\\
p_F&:=&{\Bigl|\sqrt{F}+\sqrt{(d-1)(1-F)}\Bigr|^2\over d}\ .
\label{entdp}
\end{eqnarray}
\smallskip

\noindent
{\bf Remarks 2.3}

\noindent
$(i)$\quad
The assumption of the previous proposition amounts to ask
${\cal R}(\rho_F)$ to
be regular of order $1$ with respect to the subgroup $H\subset G$ of
permutations on
$\{2,3,\cdots, d\}$.
Indeed, the leaf is $G$-invariant,
so that the $d$ states $|\phi_j\rangle=V^j|w\rangle$, $j=0,1,\ldots,d-1$,
obtained via cyclic permutations,
must be invariant under the remaining $(d-1)!$ permutations
This is possible only if $d-1$ of the $d$ components of the optimal
vector $|w\rangle$ are equal.

\noindent
$(ii)$\quad If $|w\rangle$ has three different
components, then the decompositions (\ref{ergav}) contain at least
$d(d-1)$ different terms.

\noindent
$(iii)$\quad
In section 3 we will show that, upon identification of $p_F$ with 
the quantity $\gamma(F)$ in~\cite{TV}, the entanglement of 
formation calculated
there is given by (\ref{entdp}) and (\ref{entd0}) in a range
$F^{**} \geq F > 1/d$.
The upper limit $F^{**}$ is a particular bifurcation point which was
discovered in~\cite{TV} and that will be reinterpreted accordingly within
the framework of this work.
\medskip

\noindent {\bf Proof}:\quad
By hypothesis,
$\displaystyle
\rho_F={1\over d}\sum_{j=0}^{d-1}V^j|w\rangle\langle w|V^{-j}$ is an optimal
decomposition with
entanglement
\begin{equation}
\label{opdecmin}
E(\rho_F;M_d({\bf C}),{\cal A})
=\sum_{j=1}^ds\left(|\langle j|w\rangle|^2\right)\ .
\end{equation}
Also, taking into account Remark 2.2 and 2.3, and decomposing
$$
|w\rangle=\sqrt{F}|\psi\rangle+
\varepsilon\sqrt{1-F}|w^\perp_1\rangle
=\alpha|1\rangle+\beta\sum_{j=2}^d|j\rangle\ ,
$$
where
$\varepsilon$ is a pure phase, it follows that
$|w^\perp_1\rangle=(\sqrt{d}|1\rangle-|\psi\rangle)/\sqrt{d-1}$
and
$$
|w\rangle={1\over\sqrt{d}}\Bigl[
\Bigl(\sqrt{F}+\varepsilon\sqrt{(1-F)(d-1)}\Bigr)|1\rangle+
\Bigl(\sqrt{F}-\varepsilon\sqrt{{1-F\over
d-1}}\Bigr)\sum_{j=2}^d|j\rangle\Bigr]\ .
$$
With $\xi:=2{\cal R}{\rm e}(\varepsilon)$, the right hand side
of~(\ref{opdecmin}) reads
\begin{eqnarray*}
{\cal S}(\xi)&=& s(p(\xi)) +
(d-1) s({1 - p(\xi) \over d-1}),\\
p(\xi)&=&{F+(1-F)(d-1)+\xi\sqrt{F(1-F)(d-1)}\over d}\ .
\end{eqnarray*}
It achieves
its minimum at the maximum value of $p$ that is for
$\varepsilon=1$, from which the result follows. Indeed,
as we show below, $|w\rangle$ must
be real. If remark 2.3(i) applies we always get a
local extremum. Either by direct calculation or relying
on \cite{BNU1} one concludes $\epsilon = 1$.\QED
\smallskip

We now relax the hypothesis of the previous proposition and allow for
more than one $G_0$-orbit to be optimal for the entanglement
of $\rho_F$ with respect to the subalgebra $\cal A$, that is we allow the leaf
${\cal R}(\rho_F)$ to be generated by more than one $G_0$-orbit.

\noindent
{\bf Proposition 2.4}\quad
Let ${\cal A}\subset{\cal M}=M_d({\bf C})$ be chosen as in Example 2.1.
If the density matrix $\rho_F$ is invariant with respect to the permutation
group $G$ and its entanglement with respect to $\cal A$ can be
achieved at an
optimal decompositions consisting of one
$G_0$-orbits of normalized
vector states $|w\rangle\in{\bf C}^d$, with $G_0\subset G$ the subgroup of
cyclic permutations, then we have three possibilities
\begin{itemize}
\item
$\displaystyle
|w\rangle={1\over \sqrt{d}}\sum_{k=1}^d|k\rangle$ in which case $F=1$ and
$\rho_F=|\psi\rangle\langle\psi|$;
\item
$|w\rangle$ is real with $1$ component equal to
$a_1$ and $d-1$ real components
all equal to $a_2\neq a_1$;
\item
$|w\rangle$ is real with $2$ components $a_1\neq a_3$ and $d-2$ components
all equal to $a_3$ different from both $a_1$ and $a_2$.
\end{itemize}
\smallskip

\noindent
To prove the result we need a preliminary

\noindent
{\bf Lemma 2.1}\quad
The vector $|w\rangle$ whose $G_0$-orbit is optimal can be chosen real.
\medskip

\noindent {\bf Proof}:\quad
Let $v_k$, $k=1,2,\ldots, d$, be the components of $|w\rangle$ with
respect to the
chosen orthonormal basis $\{|k\rangle\}$ and
$\displaystyle
|\psi\rangle={1\over \sqrt{d}}\sum_{k=1}^d|k\rangle$.
The assumption is that
$\displaystyle
\rho_F={1\over d}\sum_{j=0}^{d-1}V^j|w\rangle\langle w|V^{-j}$;
from normalization it follows that the components of
$|w\rangle$ must satisfy
\begin{equation}
\label{const1}
\sum_{k=1}^d|w_k|^2=1\ ,\quad
\Bigl|\sum_{k=1}^d w_k\Bigr|^2=1-\sum_{\ell\neq k=1}^{d}
w_\ell^*w_j=dF\ .
\end{equation}
Further, in order to implement optimality and achieve
$E(\rho_F;{\cal M},{\cal A})$,
we minimize
\begin{equation}
\label{const2}
{\cal S}(w,\lambda,\mu):=-\sum_{k=1}^d|w_k|^2\log|w_k|^2+
\lambda\sum_{k=1}^d|w_k|^2
+\mu\sum_{\ell\neq k}w_\ell w^*_k\ ,
\end{equation}
with Lagrange multipliers $\lambda,\mu$.
Setting
$\displaystyle v:=\sum_{k=1}^dw_k=\sqrt{dF}{\rm e}^{i\theta}$,
equating to zero the derivative of~(\ref{const2})
with respect to $w_j$ and multiplying by $w_j$ we get
$$
-|w_j|^2\log|w_j|^2+(\lambda-1)|w_k|^2+\mu(v^*\,w_j-|w_j|^2)=0\ .
$$
Therefore, the quantity $v^*w_j\mu$ and thus, after summing over $j$,
also $\mu$,
must be real, whence, necessarily $w_j={\rm e}^{i\theta}v_j$,
with $v_j\in{\bf R}$, for all $j$.
The result follows by eliminating the overall phase.
\QED
\smallskip

\noindent {\bf Proof} (of Proposition 2.4):\quad
According to the previous Lemma, we choose $|w\rangle$ real and proceed to
minimize
\begin{equation}
\label{const3}
{\cal S}(w,\lambda,\mu):=-\sum_{k=1}^d w_k^2\log w_k^2+
\lambda\sum_{k=1}^d w_k^2
+\mu\sum_{k=1}^dw_k\ .
\end{equation}
Because of convexity, the function
$g(x):=-x\log x^2$ intersects the straight line $f(x):=
2(1-\lambda)x-\mu$ in at most three points on $[-1,1]$.
Therefore, the $d$ solutions to
$$
-2w_k\log w_k^2-2w_k+2\lambda w_k+\mu=0\ ,
$$
can have at most three different real values, $a_i$, $i=1,2,3$.
We denote by $n_i$
the number of times they appear among the components and
consider the functional
\begin{equation}
\label{const4}
{\cal S}(\vec{a};\vec{n}\lambda,\mu,\nu):=-\sum_{i=1}^3n_ia_i^2
\log a_i^2+\lambda\sum_{i=1}^3n_ia_i^2+\mu\sum_{i=1}^3n_ia_i\ ,
\end{equation}
where we treat the $n_i$'s as continuous variables constrained by
$n_1+n_2+n_3=d$.
Minimizing~(\ref{const4}) yields the following equations
\begin{eqnarray}
\label{const5}
&&
n_i(a_i\log a_i^2+a_i-\lambda a_i-\mu)=0\ ,\quad i=1,2,3\\
\label{const6}
&&
-a_i^2\log a_i^2+\lambda a_i^2+\mu a_i+\nu\ ,\quad
i=1,2,3\ .
\end{eqnarray}
It follows that, if $n_i>0$, $i=1,2,3$, then,
$\sum_{i=1}^3(\mu a_i+2\nu+2a_i^2)=0$,
$i=1,2,3$, and thus $a=b=c$.
This case corresponds to $\rho_{F=1}=|\psi\rangle\langle\psi|$, a pure state,
with null entanglement with respect to $\cal A$.
Therefore, if there are three different intersections, the minimum
entanglement is reached at the boundary values of $n_i$, $i=1,2,3$, that is,
without loss of generality, at $n_1=n_2=1$ and $n_3=d-2$.
If there are two intersections, that is if, without loss of generality,
$n_3=0$ and $a_1\neq a_2=a_3$, then, from~(\ref{const5},\ref{const6}),
we calculate
$\mu=-2(a_1+a_2)$, $\mu=a_1a_2$ and deduce the equality
$$
a_1^2-a_2^2+a_1a_2\log{a_2^2\over a_1^2}=0\ .
$$
For fixed $a_1$,  because of their convexity properties,
the two functions
$\displaystyle
f(x):=\log{a_1^2\over x^2}$ and $g(x):=\displaystyle
{a_1\over x}-{x\over a_1}$ intersect at $x=a_1$, but, at no other points.
Therefore, the entanglement is again minimal at the boundary, that is at ,
say $n_1=1$ and $n_2=d-1$.
\QED
\medskip

\noindent
{\bf Remark 2.4}\quad
Lagrange multipliers have been used in \cite{TV} in order to calculate the
entanglement of formation of isotropic states of bipartite quantum systems,
where it is shown that, when $F>1/d$, the
optimal decomposers have only two different components.
We shall relate those results to ours in the following section, where we also
discuss the fact, discovered in \cite{TV}, stating there
is a bifurcation point $F^{**}$ such that
the entanglement of formation is linear in $F$ between $F^{**}$
and $F=1$.
\medskip

Proposition 2.4 shows that when the vector $|w\rangle$ has
only two different
components, then we reduce to the case discussed in Proposition 2.3.
Instead, when $|w\rangle$ has three different
components, which is possible in a range of values of $F$,
then we have more than one optimal decompositions. If $d=3$
one gets at least two.
Notice that these results are obtained under the
hypothesis that $G_0$-orbits
of vectors $|w\rangle$ provide optimal decompositions for
the entanglement
of $\rho_F$ with respect to the subalgebra ${\cal A}$.

This fact is linked to the convexity of the
function~(\ref{conc}), which, as observed in the discussion
of Case 2, fail in a neighborhood of $F=1$:  If $F \geq F^{**}$
one needs two orbits: the optimal orbit for $F = F^{**}$
and the singlet for $F=1$, just as observed in \cite{TV}.
Consequently, for $F^{**} < F < 1$ no $G_0$-orbits can
be optimal.
\bigskip

\noindent
{\bf 3. ENTANGLEMENT AND ENTANGLEMENT OF FORMATION}
\medskip

In this section we establish a one-to-one correspondence
between the results of the previous section, in particular
proposition 2.3, and the entanglement of formation of highly
symmetric states as examined in \cite{TV}. This concerns mainly
the region $(1/d) \leq F$. From \cite{TV} we learned the
existence of the bifurcation point $F^{**}$. On the other hand,
our results in the region $(1/d) < F \leq F^{**}$
can be converted into those found by Terhal and Volbrecht.
Indeed, the value of the entanglement
of formation will be proved to be just (\ref{entd0}).

To this end we consider the tensor product
${\cal M}:={\cal A}\otimes {\cal B}$ of the full
$d\times d$ matrix algebra, denoted  by $\cal A$,
with a copy, $\cal B$, of itself.
We fix an orthonormal
basis $\{|j\rangle\}$ of ${\bf C}^d$ and given any
density matrix, that is a state  on $\cal A$,
\begin{equation}
\label{dm}
\rho_{A}=\sum_{j,k}R_{jk}|j\rangle\langle k|\ ,\quad
R=[R_{jk}]\geq 0\ ,\quad {\rm Tr}R=1\ ,
\end{equation}
we embed it as $D[\rho_A]$ into the state space
of $\cal M$ according to the following

\noindent
{\bf Definition 3.1}\quad
Let $D$ be the linear map associating matrix units
$|j\rangle\langle k|$ of $\cal A$ with matrix units
$\{|j\rangle\langle k|\otimes|j\rangle\langle k|$ of $\cal M$.
We shall refer to it as the {\it doubling map}.
It transforms states
$\rho_A$ on $\cal A$ into states on ${\cal M}={\cal A}\otimes{\cal B}$ 
of the form
\begin{equation}
\label{doubling1}
\rho_{A}\longmapsto D[\rho_A]:=\sum_{j,k}R_{jk}|j\rangle\langle
k|\otimes |j\rangle\langle k|\ ,
\end{equation}
\medskip

\noindent
{\bf Remark 3.1}\quad
This yields the
class of density matrices in Example 2.3, which we shall refer
to as diagonal class (with respect to the chosen basis).
On the given diagonal class the doubling map can be inverted
\begin{equation}
\label{doubling2} D^{-1}:
\rho_{AB}=\sum_{j,k} R_{j,k} |j \rangle\langle k| \otimes
|j \rangle\langle k| \longmapsto
\rho_A= \sum_{j,k} R_{j,k} |j \rangle\langle k|\ .
\end{equation}
The argument developed in Example 2.3 ensures that decompositions
of $\rho_A$ can be mapped onto
decompositions of $D[\rho_A]$.
Vice versa, decompositions of $\rho_{AB}$ provide
decompositions for the diagonal class of $\rho_A$ by
applying $D^{-1}$.
Moreover,
if ${\cal A}_0\subset{\cal A}$ denotes the subalgebra of diagonal
matrices in the given, fixed representation, then
$S(\rho\rstr{\cal A}_{0})=S(D[\rho_A]\rstr{\cal A})$.
Therefore: {\it The entanglement is preserved by $D$,} in the sense that
\begin{equation}
\label{doubling3}
E(\rho_A;{\cal A},{\cal A}_{0})=E(D[\rho_A];{\cal A}\otimes{\cal B},
{\cal A})\ .
\end{equation}
\medskip

In~\cite{TV} the entanglement of formation has been
calculated for the isotropic states
\begin{equation}
\label{isotrst}
\omega_F={1-F\over d^2-1}({\bf 1}_{AB}-|\Psi\rangle\langle\Psi|)+
F|\Psi\rangle\langle\Psi|\ .
\end{equation}
In the above expression ${\bf 1}_{AB}$ is the identity for the algebra
${\cal A}\otimes{\cal B}$ and
\begin{equation}
\label{symmst}
|\Psi\rangle={1\over \sqrt{d}}\sum_{j=1}|j\rangle\otimes|j\rangle\ .
\end{equation}
\smallskip

\noindent
{\bf Remark 3.2}\quad
The isotropic states are invariant
under the group $\cal G$ of all unitaries of the form
$U\otimes\tilde{U}$ where $\langle
a|U|b\rangle= \langle a|\tilde{U}|b\rangle^*$,
\begin{equation}
\label{Uinv}
U\otimes\tilde{U}\omega_FU^{-1}\otimes\tilde{U}^{-1}=\omega_F\ .
\end{equation}
As in Remark 2.2, it follows that $\omega_F$ can be expressed
as the following average
with respect to the Haar measure ${\rm d}_GU$,
\begin{equation}
\label{ergav2}
\omega_F=\int_{\cal G}{\rm d}_{{\cal G}}U\,
U\otimes\tilde{U}|\Phi\rangle\langle\Phi|U^{-1}\otimes\tilde{U}^{-1}\ ,
\end{equation}
if and only if
$F=\langle\Psi|\omega_F|\Psi\rangle=|\langle\Psi|\Phi\rangle|^2$.
\medskip

We compare the isotropic state
$\omega_F$ with the doubling of $\rho_F$ in~(\ref{F1}),
\begin{eqnarray}
\nonumber
D[\rho_F]&=&{1-F\over d-1}\Bigl(D[{\bf 1}_A]
-D[|\psi\rangle\langle\psi]|\Bigr)\,+\,
FD[|\psi\rangle\langle\psi|]\\
\label{doub}
&=&{1-F\over d-1}\Bigl(\sum_{j=1}^d|j\rangle\langle
j|\otimes|j\rangle\langle j|-|\Psi\rangle\langle\Psi]|\Bigr)\,+\,
F|\Psi\rangle\langle\Psi|\ .
\end{eqnarray}

\noindent
{\bf Proposition 3.1}\quad
Let $F>1/d$ and consider the decomposition
$$
\omega_F={1\over d!}\sum_\pi
U_\pi^{-1}\otimes U_\pi^{-1}|\Phi\rangle\langle\Phi|U_\pi\otimes
U_{\pi}
$$ 
by means of the unitaries $U_\pi$ that implement the permutation group
$G$.
If the latter 
is optimal for the
entanglement of formation $E(\omega_F)$ with
$|\Phi\rangle\langle\Phi|$ in the diagonal space, then
$E(\omega_F)=E(\rho_F, {\cal A},{\cal A}_0)$.
\medskip

\noindent
{\bf Proof}:\quad
The $d!$ unitaries $U_\pi$ 
form a subgroup $G\otimes G$ of the group of unitaries
in Remark 3.2; they implement the permutation of the chosen basis
$\{|j\rangle\otimes|j\rangle\}$ of the diagonal space.
Then,
$\langle\Psi|\omega_F|\Psi\rangle=\langle\Psi|D[\rho_F]|\Psi\rangle=F$
and
$$
D[\rho_F]={1\over d!}\sum_\pi
U_\pi^{-1}\otimes U_\pi^{-1}|\Phi\rangle\langle\Phi|U_\pi\otimes
U_{\pi}\ .
$$
If $|\Phi\rangle\langle\Phi|$ is optimal for $\omega_F$, it turns out
from Proposition 2.2 that the decomposeres
$U\otimes\tilde{U}|\Phi\rangle\langle\Phi|U^{-1}\otimes\tilde{U}^{-1}$
are optimal, too.
Thus the result follows
from Proposition 2.1.
\QED

\noindent
{\bf Remarks 3.3}

\noindent
$(i)$\quad
If $F>1/d$ the isotropic state $\omega_F$ is entangled. When
$F\leq1/d$ it becomes separable. There exist
several proofs of this fact, e.g.\cite{HH}.
\medskip

\noindent $(ii)$ \quad
In view of Remark 2.3(ii), the previous
proposition establishes a link between our results and those
of~\cite{TV}. In \cite{TV} a new symmetry breaking bifurcation
point was observed at $F=8/9$ when $d=3$.
The doubling map makes it correspond
to a bifurcation point within case 2 of the previous section
at the same value of $F$
The numerical analysis in~\cite{BNU2} missed it, the needed accuracy
being of the order of $10^{-4}$.
In both cases the leaves ${\cal R}(\omega_F)$, respectively 
${\cal R}(\rho_F)$, are identical for all $F$
within $F^{**} = 8/9< F < 1$. This unique leaf is generated
by the optimal decompositions of $\omega_{8/9}$ respectively
$\rho_{8/9}$, which form one orbit, and by the pure state
$\omega_1$ given by (\ref{symmst}) respectively $\rho_1$.
The latter orbits are singlets.

\noindent
$(iii)$\quad
The entanglement of $\rho_1$ and $\rho_{8/9}$ that generate the leaf
discussed in the previous remark do not coincide,
\begin{equation}
\label{counter}
E(\rho_1;{\cal M},{\cal A})=\ln3\ ,\quad
E(\rho_{8/9};{\cal M},{\cal A})=\ln 3-{1\over 3}\ln 2\ .
\end{equation}
\medskip

We shall now relate the remark above to another observation
which again relate entanglement of different algebras
with one another. \medskip

From Case 1 in section 2, we know that vectors of the form
$\pmatrix{x\cr y}$ and $\pmatrix{y\cr x}$, with $x^2+y^2=1$ generate
the leaf of some state $\rho_2$ on $M_2({\bf C})$.
These $2$-dimensional vectors can be embedded in ${\bf C}^3$
as follows,
\begin{equation}
\label{contr1}
|w_1\rangle=\pmatrix{x\cr y/\sqrt{2}\cr y/\sqrt{2}}\ ,\quad
|w_2\rangle=\pmatrix{y\cr x/\sqrt{2}\cr x/\sqrt{2}}\ .
\end{equation}
With them we construct the density matrix in $M_3({\bf C})$ of the form
\begin{equation}
\label{contrad1}
\tilde{\rho}_3=
\lambda|w_1\rangle\langle w_1|+(1-\lambda)|w_2\rangle\langle w_2|
=\pmatrix{a&b&b\cr
b&c&c\cr
b&c&c}\ .
\end{equation}
It is easy to check that powers of $\tilde{\rho}_3$ have the same
structure which is thus inherited by $\sqrt{\tilde{\rho}_3}$ .
It thus follows that
$\sqrt{\tilde{\rho}_3}|\phi\rangle=\pmatrix{u\cr v\cr v}$ for any
$|\phi \rangle $.
The discussion of Example 2.3 assures and that the optimal decomposers of
$\tilde{\rho}_3$ with respect to the entanglement
$E(\tilde{\rho};M_3({\bf C}),{\cal A}_3)$, with ${\cal A}_3$
the maximally Abelian subalgebra in the chosen representation,
have again the same form.
But then, being $\pmatrix{x\cr y}$ and $\pmatrix{y\cr x}$ optimal
with respect to $E(\rho_2;M_2({\bf C}),{\cal A}_2)$,
(\ref{contrad1}) is itself an optimal decomposition
of $\tilde{\rho}_3$ with respect to
$E(\tilde{\rho}_3;M_3({\bf C}),{\cal A}_3)$.

According to the discussion at the beginning of this section,
it also follows that the doubling map
\begin{eqnarray}
\label{contr2a}
&&|w_1\rangle\mapsto|W_1\rangle=x|1\rangle\otimes|1\rangle
+{y\over\sqrt{2}}\Bigl(
|2\rangle\otimes|2\rangle+|3\rangle\otimes|3\rangle\Bigr)\\
\label{contr2b}
&&|w_2\rangle\mapsto|W_2\rangle=y|1\rangle\otimes|1\rangle
+{x\over\sqrt{2}}\Bigl(
|2\rangle\otimes|2\rangle+|3\rangle\otimes|3\rangle\Bigr)\ ,
\end{eqnarray}
provides optimal decomposers, too. In particular, for given
$x, y$ on the unit circle the pure states
$|W_j \rangle\langle W_j|$, $j=1,2$, generate a leaf of the
entanglement of formation functional on which it is
convexly linear.

Moreover, for $x=1/\sqrt{3}$ and $y=\sqrt{2/3}$, we get
$|W_1\rangle=|\Psi\rangle$, with fidelity
$F=|\langle\Psi|W_1\rangle|^{2}=1$, and
$|W_2\rangle=|\Phi_{8/9}\rangle$
with fidelity $F=|\langle\Psi|W_2\rangle|^{2}=8/9$,
indicating a reason for the bifurcation value
$F= 8/9$.

One observes that (\ref{contr2a}) and (\ref{contr2b})
become identical for $x=y=1/\sqrt{2}$ so that the doubling map gets
the vector
\begin{equation}
\label{cntr3}
|W_3\rangle={1\over\sqrt{2}}|1\rangle\otimes|1\rangle
+{1\over 2}\Bigl(|2\rangle\otimes|2\rangle+
|3\rangle\otimes|3\rangle\Bigr)\ ,
\end{equation}
which has fidelity
\begin{equation}
\label{contr4}
F=|\langle\Psi|W_3\rangle|^2={1\over 2}+
\sqrt{{2\over 3}}=p+(1-p){8\over 9}\ ,\quad
0< p=3\sqrt{6}-{7\over 2}<1\ .
\end{equation}

\noindent
Let us now consider the state
\begin{equation}
\label{counter2}
\rho_F=p|\Psi\rangle\langle\Psi|+(1-p)|\Phi_{8/9}\rangle\langle\Phi_{8/9}|
\ .
\end{equation}
By using~(\ref{counter}), it can be shown that its entanglement
$E(\rho_F)$ is larger than
$pE(\rho(1))+(1-p)E(\rho(8/9))$ for $0 < p < 1$. This implies
that convexity of  ${\cal S}(F)$ in~(\ref{opdecmin}) is lost
for $F > F^{**}$ in accordance with the discussion above.

We finally note that one can extend (\ref{contr1})
to all dimensions larger than two. Indeed, let
$z_1, z_2$ denote the components of a unit vector in
two dimensions. By similar arguments one proves that
the leaves of case 1 of the previous section are mapped
onto certain leaves belonging to the entanglement of
formation in $d+1$ dimensions by the embeddings
\begin{equation} \label{add1}
\pmatrix{z_1 \cr z_2} \, \longrightarrow \,
z_1 |00\rangle + (z_2 / \sqrt{d}) \sum_{j=2}^{d+1} |jj\rangle
\end{equation}
In particular, the embeddings of $\{z_1, z_2\}$ and
$\{z_2^*, z_1^*\}$ form an optimal pair with respect
to the entanglement of formation. One further observes
in the special case $z_1 = 1 / \sqrt{d+1}$ the embeddings
(\ref{add1}) are the totally symmetric vector
$\Psi$ in $d+1$ dimensions and
\begin{equation} \label{add2}
\sqrt{{d \over d+1}} |11\rangle + \sqrt{{1 \over d(d+1) }}
\sum_{j=2}^{d+1} |jj\rangle
\end{equation}
Its fidelity reads $F = 4d/(1+d)^2$, and we see as above
\begin{equation} \label{add3}
F^{**}_{d+1} = 4 d (d+1)^{-2}
\end{equation}
i.~e. the bifurcation value given in \cite{TV}
for $d+1 > 2$.
\bigskip

\noindent
{\bf 4. CONCLUSIONS}

We have studied in several examples the entanglement
defined by a maximal
commuting subalgebra of a full matrix algebra, and
in its relation to the entanglement of formation.
Apart from
its actual numerical value, what is interesting is the
structure of both entanglement functionals upon
the space of states, and their separation into
different leaves. To some extent these leaves can be found
by applying group theoretical considerations. They show a
rich structure with varying stability under the groups
under consideration, Since the same
group appears in different algebraic contexts, it can be
shown that the decompositions of states on different
algebras can be related. This helps to control the optimal
decompositions and to understand their variety. This new
technique is shown at work in several examples:
The doubling map relates two quite
different lines of research which had been considered
almost independently up to now. In particular we have a further
proof of the entanglement of formation results for
isotropic states of Terhal and Volbrecht in the
region $(1/n) \leq F \leq F^{**}$, \cite{TV}.
Another embedding map
verifies their bifurcation point $F^{**}$ close to $F=1$
as a footprint of a symmetry-breaking in two dimensions.
It belongs to class of maps which change entanglement but
not the leaves. The leaves are respected because the
entanglements differ just by a convexly linear function.

It should be clear that we only provide some distinguished
first examples of our embedding procedures which can
connect various entanglement problems and, evidently, other
ones which are defined via convex or concave roofs, for
example general entanglement monotones or Holevo
(1-shot) capacities.

\end{document}